\documentclass[conference]{IEEEtran}
\usepackage{verbatim}
\usepackage[hidelinks]{hyperref}
\usepackage{float} 
\usepackage{booktabs} % For formal tables
\usepackage{tabularx} % For adjustable-width columns
\usepackage{caption}  % For custom captions
\usepackage{subcaption} % For subfigures
\usepackage{multirow} % in preamble
\usepackage{cite}
\usepackage{amsmath,amssymb,amsfonts}
\usepackage{authblk}
\usepackage{graphicx,color}
\usepackage{textcomp}
\usepackage{xcolor}
\usepackage{algorithm,algorithmic}

\hypersetup{hidelinks}

\def\BibTeX{{\rm B\kern-.05em{\sc i\kern-.025em b}\kern-.08em
    T\kern-.1667em\lower.7ex\hbox{E}\kern-.125emX}}
\AtBeginDocument{\definecolor{tmlcncolor}{cmyk}{0.93,0.59,0.15,0.02}\definecolor{NavyBlue}{RGB}{0,86,125}}

\begin{document}

\title{TDoA-Based Self-Supervised Channel Charting with NLoS Mitigation\thanks{This work has been submitted to the IEEE for possible publication. Copyright may be transferred without notice, after which this version may no longer be accessible.}}

\author[1]{Mohsen Ahadi\thanks{Corresponding author: mohsen.ahadi@eurecom.fr}}
\author[1]{Omid Esrafilian}
\author[1]{Florian Kaltenberger}
\author[2]{Adeel Malik}

\affil[1]{Communication Systems Department, EURECOM, Sophia Antipolis, France}
\affil[2]{Firecell, Nice, France}
\date{}

\maketitle

\begin{center}
    \small
    © 2026 The Authors. This work is licensed under a Creative Commons Attribution 4.0 License (CC-BY).
    
    The work has also been published by IEEE at: \texttt{https://doi.org/10.1109/TMLCN.2026.3688619}
\end{center}

\begin{abstract}
Channel Charting (CC) has emerged as a promising framework for data-driven radio localization, yet existing approaches often struggle to scale globally and to handle the distortions introduced by non-line-of-sight (NLoS) conditions. In this work, we propose a novel CC method that leverages Channel Impulse Response (CIR) data enriched with practical features such as Time Difference of Arrival (TDoA) and Transmission Reception Point (TRP) locations, enabling a TDoA-based self-supervised localization function on a global scale. The proposed framework is further enhanced with short-interval User Equipment (UE) displacement measurements, which improve the continuity and robustness of the learned positioning function. Our algorithm incorporates a mechanism to identify and mask NLoS-induced noisy measurements, leading to significant performance gains. We present the evaluations of our proposed models in a real 5G testbed and benchmarked against centimeter-accurate Real-Time Kinematic (RTK) positioning, in an O-RAN–based 5G network by OpenAirInterface (OAI) software at EURECOM. It demonstrates results that outperform the state-of-the-art semi-supervised and self-supervised CC approaches in a real-world scenario. The results show localization accuracies of 2–4 meters in 90\% of cases, across varying NLoS ratios. Furthermore, we provide public datasets of CIR recordings, along with the true position labels used in this paper's evaluation.
\end{abstract}

\section*{Keywords}
5G, positioning, O-RAN, channel charting, OAI, RTK


\markboth{Ahadi \MakeLowercase{\textit{et al.}}: TDoA-Based Self-Supervised Channel Charting with NLoS Mitigation}{}

\section{Introduction}
Standard 5G New Radio (5G NR) methods for positioning a User Equipment (UE) in a wireless networks rely on estimating channel parameters such as Received Signal Strength Indicator (RSSI), Time of Arrival (ToA), Time Difference of Arrival (TDoA), and Angle of Arrival/Departure (AoA/AoD)~\cite{3gpp38.305,10119056}. These measurements enable triangulation or trilateration of a device’s position, while Channel State Information (CSI) provides a richer representation by capturing detailed channel properties and environmental influences. CSI is particularly valuable in challenging conditions, such as multipath propagation and Non-Line-of-Sight (NLoS) scenarios common in dense urban or indoor environments.

Direct CSI-based positioning methods are broadly categorized into supervised and self-supervised approaches~\cite{9264122}. Supervised learning, especially fingerprinting, leverages a database of signal features (e.g., CSI or Multi-Path Components, MPC) collected at known locations to train a model that can later predict positions. This works well in environments with relatively stable channel characteristics and if a large dataset with labels is available. In contrast, self-supervised methods bypass labeled datasets by directly inferring position from collected data through loss functions, making them more adaptable to dynamic environments where maintaining large labeled datasets is impractical.

While CSI offers comprehensive insights into wireless propagation, in Orthogonal Frequency Division Multiplexing (OFDM) systems, it represents high-dimensional data, posing challenges for direct use in positioning tasks. Channel Charting (CC) is a dimensionality reduction method that addresses this by learning a map of the wireless environment from CSI or Channel Impulse Response (CIR) to a lower dimensional embedding space, enabling device localization and tracking even in complex propagation scenarios~\cite{ferrand2023wireless}. The CC goal is to transform data from a space of dimension $D$ to a lower dimension $d \ll D$, while preserving the physical relationships of the original space in the embedding space.

Non-parametric CC methods such as Multidimensional Scaling (MDS), Isometric Mapping (ISOMAP), and Principal Component Analysis (PCA) have shown effectiveness in reducing dimensionality while retaining key structural information~\cite{van2009dimensionality}. However, these methods are unable to generalize a mapping function to handle unseen data without recomputation. This limits their applicability in dynamic scenarios. Parametric approaches, on the other hand, particularly deep learning-based methods, overcome this limitation by learning mappings that generalize to unseen inputs, offering a significant advantage for practical deployment.
Recent advances in parametric methods, also called self-supervised CC with deep metric learning~\cite{8444621,8645281,9109875,Huang2019ImprovingCC,9771913,8919897,9448128,9833925,10070385,10074200,euchner2023augmenting,stahlke2023velocitybased,taner2023channel} (see Section~\ref{sec:Related work}) have pushed the field forward. Nonetheless, these methods still fall short of the accuracy achieved by supervised fingerprinting or classical triangulation approaches, even under line-of-sight (LoS) conditions.

In our earlier work~\cite{11101632}, we introduced a CC method with sensor fusion that outperformed state-of-the-art CC approaches and classical TDoA baselines under LoS conditions. However, the results were obtained in a simulated environment and relied on ToA availability and predominantly LoS propagation. The present work extends~\cite{11101632} toward a practical 5G UL-TDoA positioning setting. Specifically, we (i) replace ToA assumptions with a UL-TDoA formulation that does not require RAN-UE synchronization, (ii) incorporate known TRP locations to directly learn a global mapping in a common Cartesian coordinate system, (iii) introduce a lightweight CIR-peak-power-based masking mechanism to mitigate NLoS-induced distortions, and (iv) integrate short-interval displacement constraints to improve trajectory continuity in low-LoS measurements.

The performance of the proposed CC is evaluated first by synthetic data from the Matlab Ray-Tracing environment, and secondly by a real-world positioning setup called "GEO-5G testbed" at EURECOM. For this, we deployed an O-RAN 5G network using OpenAirInterface (OAI) and tested our new CC model. We will discuss how hardware impairments and practical considerations in our testbed affect the position estimation and how we manage to solve them.
Our CC predictions are compared with the Real-Time Kinematic (RTK) positioning, which is an enhanced solution to Global Positioning System (GPS)~\cite{Wang2020GPSBDSRtk}. The main contributions of this work are:
\begin{itemize}
    \item A novel CC method that predicts UE's position in mixed LoS/NLoS conditions with 2-4 m accuracy,
    \item A practical approach by incorporating CIR, TDoA measurements, TRP locations, and displacement measurements to achieve a TDoA-based self-supervised and global-scale position estimation in mobile scenarios,
    \item Evaluating the performance in both simulation and real 5G O-RAN testbed,
    \item Publicly releasing the 5G CIR dataset of the UE's trajectory as well as ground truth positions. \footnote{\url{https://ieee-dataport.org/documents/eurecom-5g-srs-dataset}}
\end{itemize}

The remainder of this paper is organized as follows. Section~\ref{sec:Related work} reviews the related work. Section~\ref{sec:sysmodel} introduces the system model. Section~\ref{sec:CCtdoa} presents the proposed CC approach based on TDoA knowledge and sensor fusion with NLoS mitigation. The evaluations and results are discussed in Section~\ref{sec:results}, and Section~\ref{sec:conclusion} concludes the paper and outlines directions for future work.

\section{Related Work}
\label{sec:Related work}

Time-based localization methods such as ToA and TDoA are well known to suffer performance degradation in mixed LoS/NLoS environments due to bias introduced by path delays under NLoS propagation. A large body of work therefore investigates NLoS identification and mitigation using only timing measurements and geometric consistency. For instance, residual analysis and subset consistency tests have been used to detect and exclude NLoS links by comparing positioning residuals across different anchor subsets \cite{Guvenc2009Survey, Hua2018ResidualNLOS}. Robust optimization formulations that minimize outlier influence in TDoA systems have also been proposed, including $\ell_1$-norm and outlier-aware cost functions to reduce the impact of NLoS biases without requiring received signal strength or power measurements \cite{Picard2012OutliersTDOA, Ma2019OutlierTDOA}. Survey articles summarize these approaches and related trade-offs in terms of prior information and complexity \cite{Nkrow2022NLOSsurvey}.
In contrast to classical time-only methods, our CC-based approach can additionally exploit CIR-derived statistics for lightweight NLoS masking and combine it with geometry-based filtering to discard noisy or NLoS-corrupted ToA/TDoA measurements, while still grounding localization in TDoA geometry.

CC for localization in wireless networks has been used for the first time in~\cite{8444621} from a single base station (BS) with multiple antennas, and in~\cite{8645281,9109875} from multiple massive MIMO BSs in space. Since CC relies on dimensionality reduction of the CSI,~\cite{Huang2019ImprovingCC} and~\cite{9771913} used autoencoders to improve this task. 

A Siamese neural network is proposed in~\cite{8919897} and~\cite{10070385} that takes random pairs of CSI first to learn a local channel chart and then transforms it into the global form using a subset of labeled data as reference points in a semi-supervised manner. In this method, the Euclidean distances of the Channel Impulse Response (CIR) measurements are used as a dissimilarity metric. To overcome the limitations of the Siamese loss function with a Euclidean distance metric, a triplet-based loss is used in~\cite{9448128,9833925} to learn the similarity between triplets of CSI data based on the distance of other side information, such as the relative recording timestamps. 

The authors of~\cite{10074200,euchner2023augmenting} combined CC with the classical localization approaches, taking ToA and AoA measurements to improve the global channel chart. Although the CSI measurements can contain rich information, none of these CC studies exploiting only CSI data have surpassed the performance of traditional triangulation-based methods, even when LoS conditions are present. In~\cite{stahlke2023velocitybased}, velocity estimation and topological map data are used for the global transformation of the CC. However, the global consistency of this algorithm relies on the length of the trajectory taken by the user. Also, the map-matching algorithm in this study works only if a unique match of the channel chart exists in the map.
Finally, in~\cite{taner2023channel}, by proposing a loss function containing a bilateration loss including multiple BSs with known locations and a triplet loss, a self-supervised CC is made in real-world coordinates. 

Motivated by the literature, in this paper, we extended our last CC with a sensor fusion algorithm in~\cite{11101632} to a more practical model. In this paper's algorithm, instead of ToA, we incorporate TDoA measurements as well as the TRP locations into the CC loss function to map CIR data into 2D UE locations. TDoA measurement is easier to obtain as the network only requires tight synchronization among RAN and not with the UE. Moreover, we show that sensor fusion can enhance the performance of CC in scenarios with limited LoS availability.
\begin{figure}[htbp]
    \centering
    \includegraphics[width=0.99\linewidth]{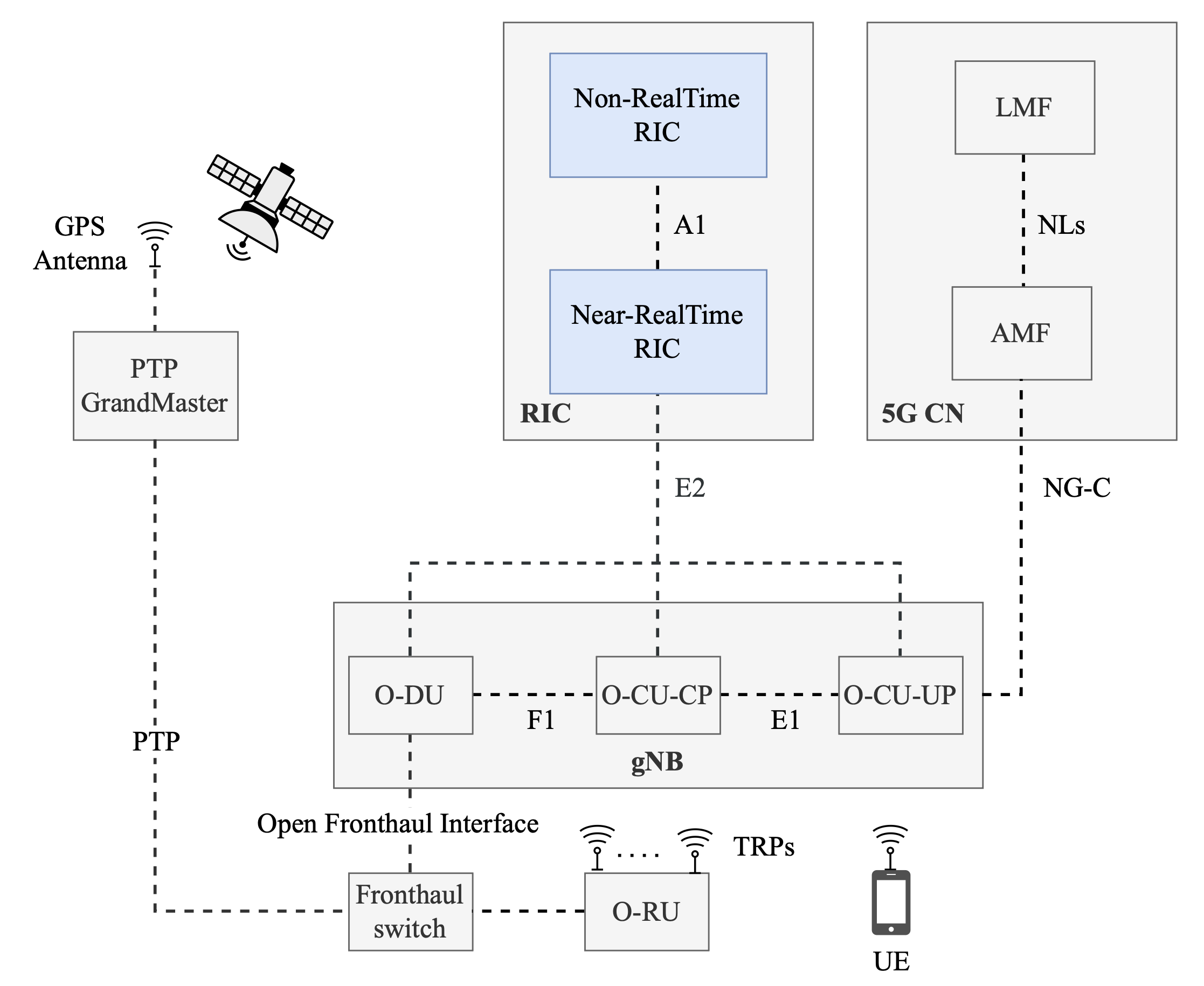}
    \caption{5G AI/ML Positioning system model. \label{fig:cc_systemmodel}}
\end{figure}

\section{System Model}\label{sec:sysmodel}
\subsection{O-RAN Network Architecture}
Figure~\ref{fig:cc_systemmodel} shows a system model for our AI/ML 5G positioning solution where we deploy an O-RAN network architecture with 7.2 functional split~\cite{oran_arch}. In this framework, the gNB is disaggregated into Open Central Unit (O-CU), Open Distributed Unit (O-DU), and Open Radio Unit (O-RU). For localizing the UE in this framework, we consider two phases of training and testing of AI/ML models over the Non-Real Time RAN Intelligent Controller (RIC) and Near-Real Time RIC, respectively. The required measurements for AI/ML models are carried over the E2 interface from O-CU-DU to RIC.

For both phases, we assume a mobile UE that transmits Up Link Sounding Reference Signals (UL-SRS) to all TRPs on RUs $k \in \{1,...,K\}$ of a gNB (while this could be generalized to multiple gNBs). Let the TRPs of the $k$-th RU have indices $m_k \in \{1,\dots,M_k\}$, with fixed and known locations $\mathbf{x}_{m_k} \in \mathbb{R}^3$ distributed in the environment. We measure the SRS channel at time instants $t\in \{1,..., T\}$. To achieve an accurate position estimation based on time measurements, a Precision Time Protocol (PTP) grandmaster provides a synchronization signal that is distributed to the RAN over a switch.  The UE follows a trajectory, and its location at each transmission time is assumed unknown and denoted by $\textbf{u}_t \in \mathbb{R}^2$, while its height is assumed fixed. 
The UE and RUs operate in an OFDM system with a total of $N_{\text{fft}}$ sub-carriers. The estimated Channel Frequency Response (CFR) of the link between the $m_k$-th TRP of $k$-th RU, and the UE at time step $t$ over all sub-carriers is denoted by $\textbf{w}_{k,m_k, t} \in \mathbb{C}^{N_{\text{fft}}}$.
The RU obtains the CIR 
$\mathbf{h}_{k,m_k,t}\in\mathbb{C}^{N_{\text{fft}}}$ 
by applying an Inverse Discrete Fourier Transform (IDFT) 
to the corresponding CFR. Collecting all $M_k$ TRPs of RU $k$ yields the RU-level CIR matrix
\begin{equation}
\mathbf{H}_{k,t} = 
\begin{bmatrix} 
\mathbf{h}_{k,1,t} \\ 
\mathbf{h}_{k,2,t} \\ 
\vdots \\ 
\mathbf{h}_{k,M_k,t}
\end{bmatrix}
\in \mathbb{C}^{M_k \times N_{\text{fft}}},
\end{equation}

Similarly, a global CIR matrix of all RUs is defined as
\begin{equation}
  \mathbf{H}_{t} = 
  \begin{bmatrix} 
  \mathbf{H}_{1,t} \\ 
  \vdots \\ 
  \mathbf{H}_{K,t}
  \end{bmatrix}
  \in \mathbb{C}^{M \times N_{\text{fft}}},
\end{equation}
where $M=\sum_{k=1}^K M_k$ is the total number of TRPs across all RUs. $\mathbf{H}_{t}$ has to be carried from CU-DU to the Near-Real Time RIC over E2 for further processing and position inference.

Moreover, only during the training phase, we assume that the UE’s displacement $d_{i,j}$ between times $t_i$ and $t_j$ is available (e.g., measured using an onboard sensor such as an IMU, laser odometry, or LiDAR). We model this displacement as
\begin{equation}\label{eq:disp}
    d_{i,j} = \| \mathbf{u}_{t_j} - \mathbf{u}_{t_i}\| + b_{i,j} + w,
\end{equation}
where $w$ denotes zero-mean Gaussian noise and $b_{i,j}$ represents a slowly varying bias term that accounts for systematic sensor drift. While the variance of $w$ typically grows with the elapsed interval $|t_j - t_i|$, the bias term $b_{i,j}$ also accumulates over time, leading to an increasing deviation between the true and measured displacement. To mitigate this drift, displacement data are collected only within a maximum interval $\epsilon$, such that $|t_j - t_i|\leq \epsilon$.

\begin{comment}
Let $\mathbf{v}(t)$ denote the user velocity at time $t$. 
The displacement between two timestamps $t_i$ and $t_j$ ,in the presence of a velocity bias $\mathbf{b}$ and noise $\mathbf{w}(t)$ is estimated as
\begin{equation}\label{eq:velocity_bais}
\begin{aligned}
    \hat{d}_{i,j} &= \int_{t_i}^{t_j} \big( \mathbf{v}(t) + \mathbf{b} + \mathbf{w}(t) \big)\, dt \\
                  &= d_{i,j} + \mathbf{b}(t_j - t_i) + \int_{t_i}^{t_j}\mathbf{w}(t)\, dt .
\end{aligned}
\end{equation}

In discrete time with sampling interval $\Delta t$, this is approximated by
\begin{equation}
    \hat{d}_{i,j} \;\approx\; \sum_{t=i}^{j-1} \big(\mathbf{v}[t] + \mathbf{b} + \mathbf{w}[t]\big)\Delta t .
\end{equation}
\end{comment}
It is important to note that the algorithm operates in a fully self-supervised manner during both the training and testing phases. Consequently, no ground-truth position label needs to be associated with the CIR measurements for either position estimation or coordinate alignment. The only distinction between the two phases lies in the input data; while displacement information is assumed to be available and leveraged during training, the model is able to generalize and function without displacement data during testing.

\subsection{Data Pre-Processing and Feature Extraction}

To enhance the robustness of the learning model, we apply several pre-processing steps to the measurements before feature extraction. As the first step in the pipeline, it is essential to apply an IFFT shifting as well as the IDFT to the CFR to reposition the zero-frequency component at the center of the time-domain CIR output. Without this step, due to hardware-specific FFT implementations and synchronization signal drift over time, the most significant peak of the CIR may appear wrapped around the beginning or the end of the CIR vector.
This leads to inconsistent peak estimation over time and across RUs.

Therefore, for each RU independently, we align the CIRs across TRPs using a peak-based procedure, ensuring that the TDoA information are preserved.
Let $\tau_{k,m,t}^{\text{peak}}$ denote the peak index of the CIR magnitude on RU $k$, TRP $m_k$, at time $t$, which is equivalent to the relative ToA at that TRP.

\begin{equation}
\tau_{k,m_k,t}^{\text{peak}} 
= \arg\max_{n \in \{1,\ldots,N_{\text{fft}}\}} 
\left| \mathbf{h}_{m_k,t}[n] \right|,
\end{equation}
where $|\cdot|$ denotes the element-wise absolute value operation.  
In real-world setups with synchronization impairments, an additional pre-processing step is required for outlier peak detection. This is achieved by applying valid ToA and TDoA bounds derived from the geometry of the testing area, similar to the filters proposed in our earlier work~\cite{ahadi2025experimentalinsightsopenairinterface5g}. It is important to note that such synchronization impairments and outliers are not typically present in simulations or in every testbed scenario.
A shifting index is selected as the earliest peak among TRPs of each RU by.

\begin{equation}
    \eta_{k,t}^{\text{shift}} 
    = \min\limits_{m \in \{1,\dots,M_k\}} \tau_{k,m_k,t}^{\text{peak}}.
\end{equation}

The following operation shifts the vector to the left and fills the tail with zeros:
\begin{equation}
    \mathbf{h}_{k,m_k,t}^{\text{shifted}}[j] = 
\begin{cases}
\mathbf{h}_{k,m_k,t}[j + \eta_{k,t}^{\text{shift}}] & \text{if } j + \eta_{k,t}^{\text{shift}} < N_{\text{fft}}, \\
0 & \text{otherwise},
\end{cases}
\end{equation}
resulting in a TDoA-aligned RU-level CIR matrix \({\mathbf{H}}_{k,t}^{\text{shifted}}\).

\begin{figure}[htbp]
    \centering
    \includegraphics[width=0.99\linewidth]{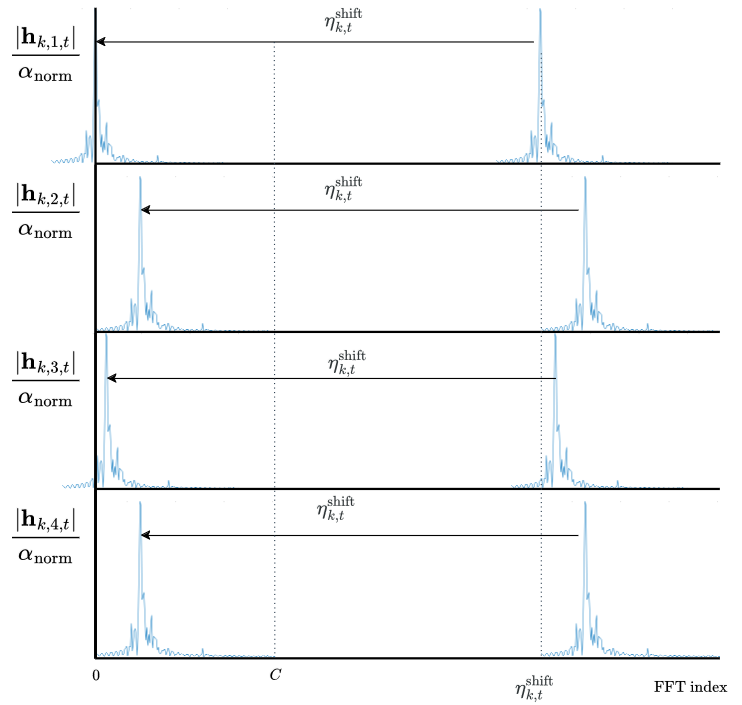}
    \caption{CIR shifting, truncation, and normalization in pre-processing while keeping the TDoA integrity}
    \label{fig:tdoa_alignment}
\end{figure}
This procedure, during both training and testing phases, ensures that the CIRs are temporally aligned across TRPs of the same RU, thereby respecting the assumption of tight time synchronization among them and the correctness of TDoA.

\begin{figure*}[htbp]
    \centering
    \includegraphics[width=0.9\linewidth]{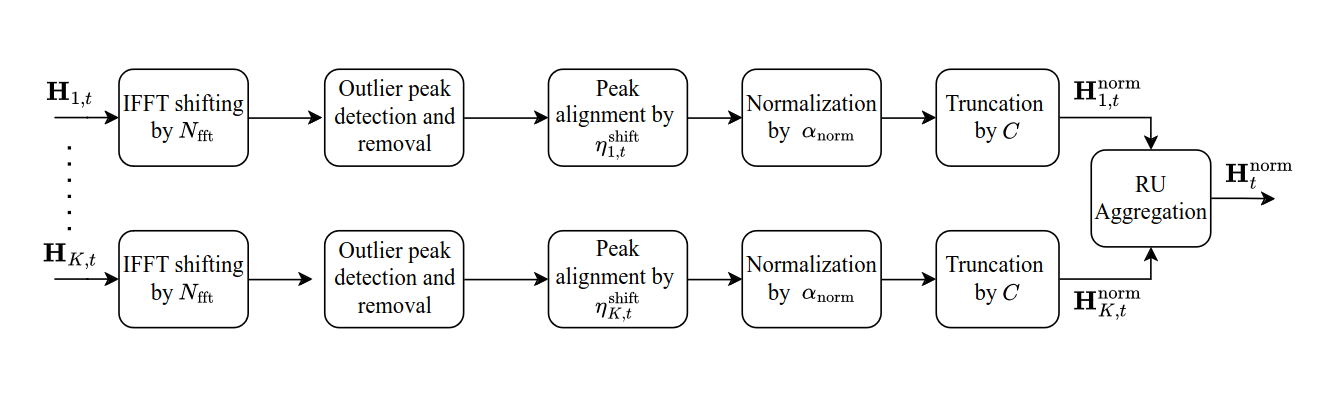}
    \caption{CIR pre-processing diagram}
    \label{fig:preprocessing_diagram}
\end{figure*}

The TDoA vector is then obtained by 
subtracting a $\tau_{k,m_{\text{ref},k},t}^{\text{peak}}$ from all $M_k$ TRPs $\tau_{k,m_k,t}^{\text{peak}}$;

\begin{equation}
\mathbf{\Delta\tau}_{k,t} 
= \Big[ \, \tau_{k,1,t}^{\text{peak}} - \tau_{k,m_{\text{ref},k},t}^{\text{peak}}, \;
              \ldots, \;
              \tau_{k,M_k,t}^{\text{peak}} - \tau_{k,m_{\text{ref},k},t}^{\text{peak}} \, \Big],
\end{equation}
where $\tau_{k,m_{\text{ref},k},t}^{\text{peak}}$ denotes the peak index at a predefined reference TRP $m_{\text{ref},k}$ on each RU. Unlike the classical TDoA method, where all TDoAs have the same reference, our approach in selecting a per-RU reference $m_{\text{ref},k}$ is to mitigate hardware synchronization impairments in real-world testing. This is further discussed in Section~\ref{sec:geo5gtestbed}.
The global CIR shifted matrix \({\mathbf{H}}_{t}^{\text{shifted}}\) is further normalized by the maximum peak magnitude observed across the entire training dataset.
Also, to form a unified input dimension for model training and testing, CIRs are truncated to only contain the first $C$ FFT indices:
\begin{equation}\label{eq:normalized}
\mathbf{H}_{t}^{\text{norm}} 
= \frac{1}{\alpha_{\text{norm}}}
\begin{bmatrix}
|\mathbf{h}_{1,t}^{\text{shifted}}| \\[4pt] 
|\mathbf{h}_{2,t}^{\text{shifted}}| \\[2pt] 
\vdots \\[2pt] 
|\mathbf{h}_{M,t}^{\text{shifted}}|
\end{bmatrix}
\in \mathbb{R}^{M \times C},
\end{equation}
where the normalization factor \(\alpha_{\text{norm}} = \max\limits_{k,m_k,t} ({ |\mathbf{h}}_{k,m_k,t}^{\text{shifted}}|)\) is computed from the training data and reused during testing to ensure consistency.
In the following sections, we leverage the training dataset
\[
\mathcal{D}_{\mathrm{tr}} 
= \big\{ \big(\boldsymbol{H}^{\mathrm{norm}}, \, \boldsymbol{\Delta{\tau}}, \, \boldsymbol{X}, \, \boldsymbol{D},\boldsymbol{T}\big) \big\},
\]
where 
$\boldsymbol{H}^{\mathrm{norm}}$ denotes the normalized global CIR matrix dataset, 
$\boldsymbol{\Delta{\tau}} \in \mathbb{R}^{M-K}$ is the vector of observed TDoA measurements across all TRPs, 
$\boldsymbol{X} \in \mathbb{R}^{M \times 3}$ contains the all TRP positions, 
$\boldsymbol{D}$ is the set of displacement measurements derived from UE's sensors during training, and $\boldsymbol{T}$ timestamps of the measurements. 
This dataset is used to learn a spatial mapping function that captures the geometric features of the environment. A summary of the pre-processing is depicted in Figure~\ref{fig:preprocessing_diagram}.
\begin{figure}[htbp]
    \centering
    \includegraphics[width=0.8\linewidth]{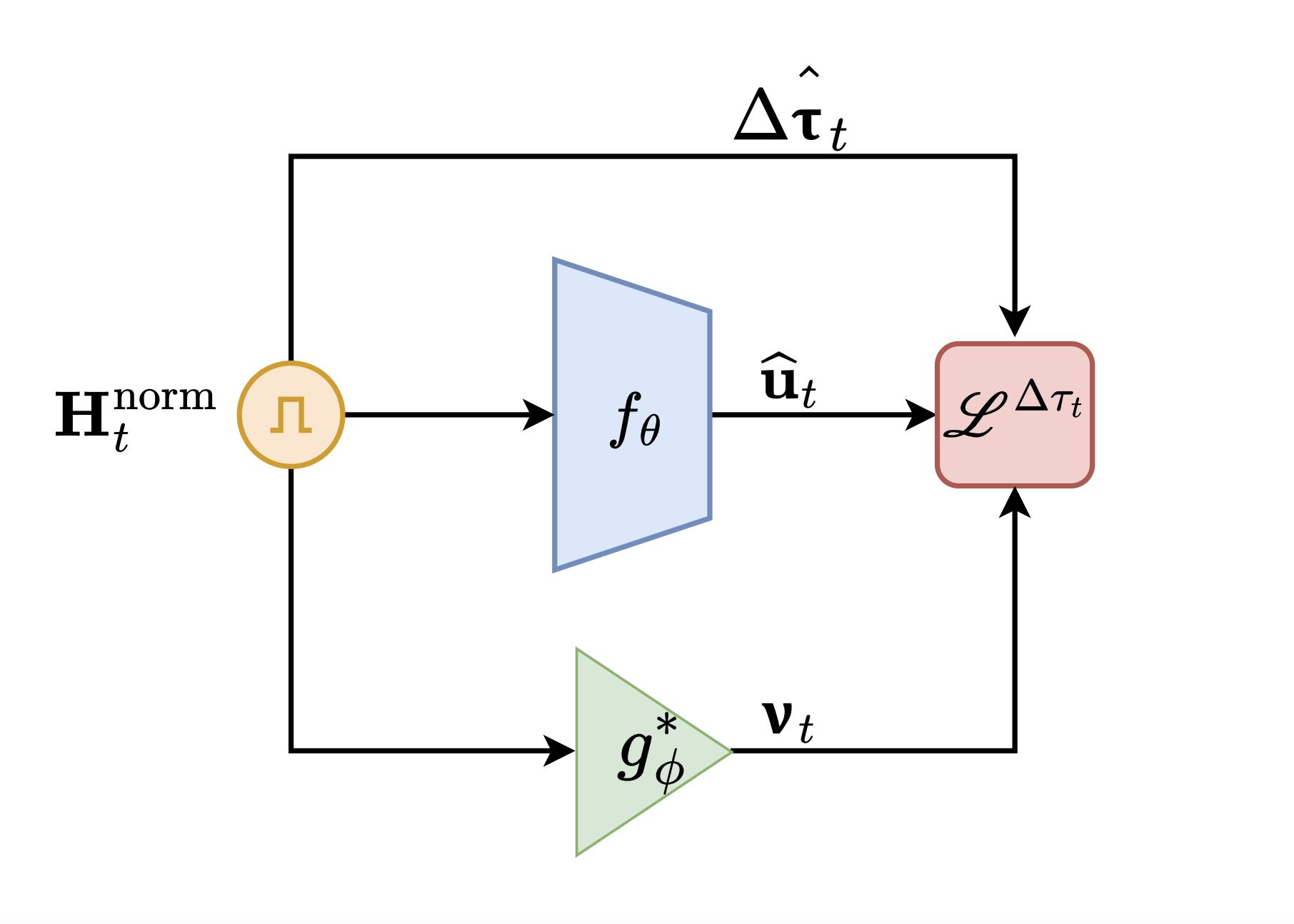}
    \caption{CC training with CIR and TDoA}
    \label{fig:tdoa_training}
\end{figure}
\begin{figure}[htbp]
    \centering
    \includegraphics[width=0.8\linewidth]{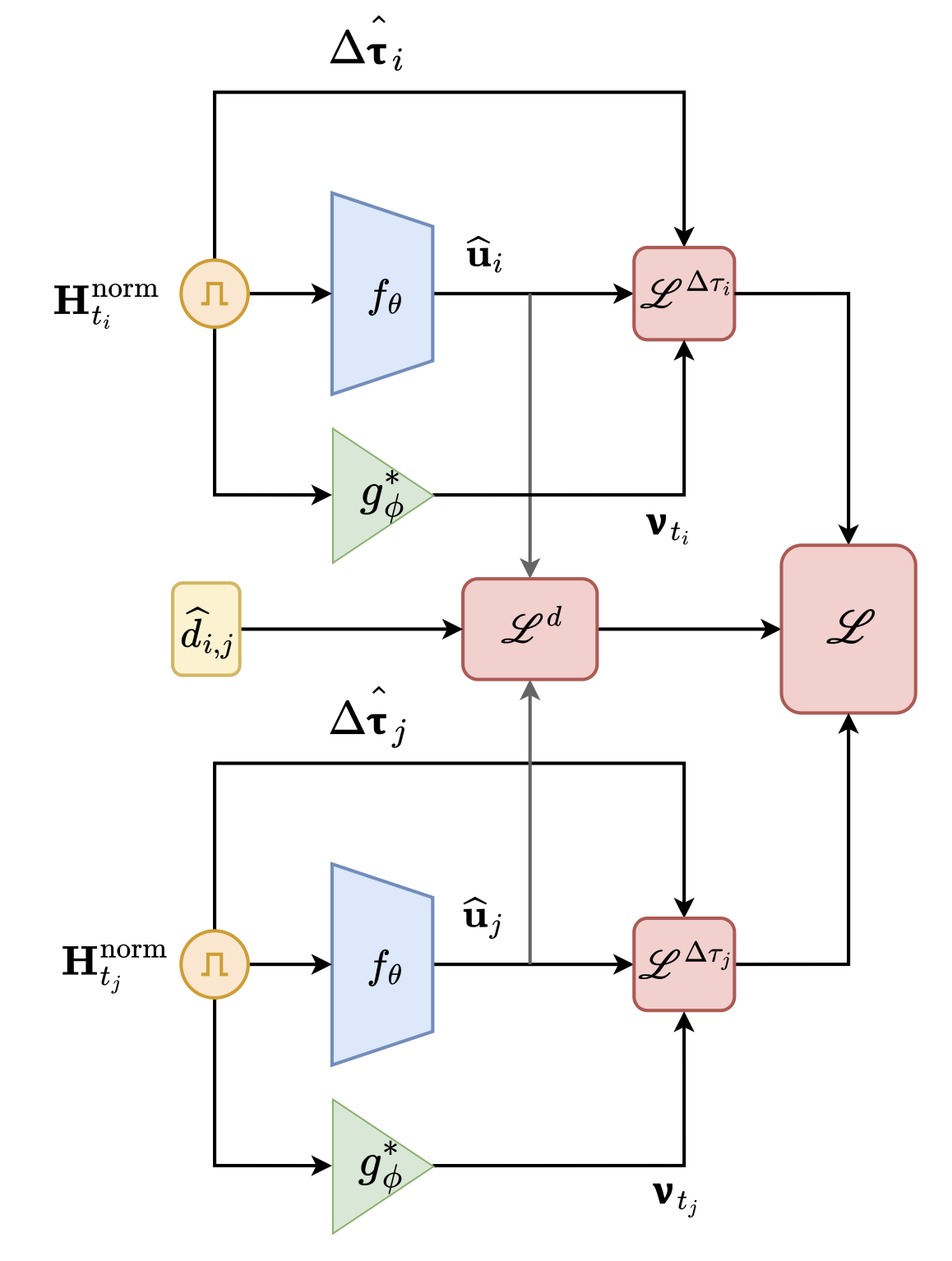}
    \caption{CC training with CIR and TDoA+displacement}
    \label{fig:tdoa_disp_training}
\end{figure}
\begin{figure}[htbp]
    \centering
    \includegraphics[width=0.8\linewidth]{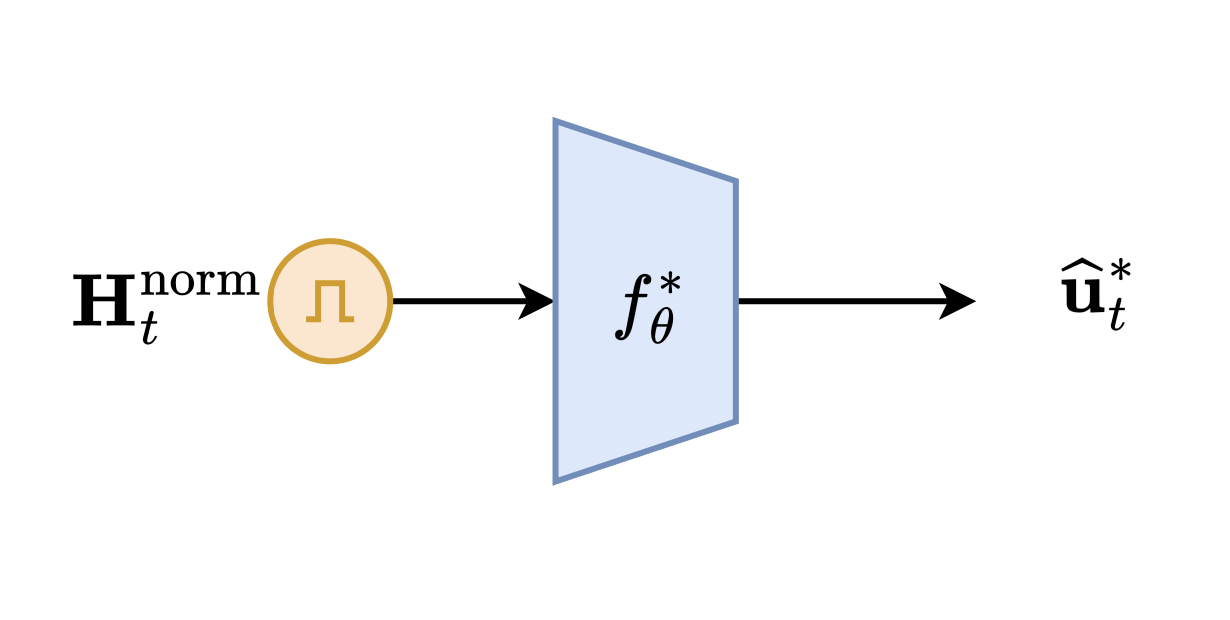}
    \caption{CC testing}
    \label{fig:CCtesting}
\end{figure}

\section{TDoA Channel Charting with Sensor Fusion and NLoS Mitigation} \label{sec:CCtdoa}
\subsection{Training Channel Charting Models}
In this section, we seek to learn a self-supervised CC function $f_{\boldsymbol{\theta}}(.)$ that transforms each CIR matrix $\mathbf{H}_{t}^{\text{norm}}$ into a $2$-dimensional embedding; 
\[
f_{\boldsymbol{\theta}} : \mathbb{R}^{M \times C} \rightarrow \mathbb{R}^{2}.
\]
The mapping is learned such that similarities in the radio channel domain (i.e., channel similarity induced by spatial proximity in the physical environment) are preserved in the embedded space, given our training dataset $\mathcal{D}_{\text{tr}}$. Deep Neural Networks (DNNs)
have proven to be powerful nonlinear function approximation and effective tools for dimensionality reduction, making them particularly well suited for learning the  $f_{\boldsymbol{\theta}}$ mapping ~\cite{8444621,8645281,9109875,Huang2019ImprovingCC,9771913,8919897,10070385,9448128,9833925,10074200,euchner2023augmenting,stahlke2023velocitybased,taner2023channel}. Therefore, we employ an embedding DNN model with the architecture described in Table~\ref{tab:embedding_model_summary}. The embedding model consists of two convolutional layers for feature extraction, followed by a flattening stage and three fully connected layers. The network gradually reduces the feature dimensionality and projects the CIR input into a compact two-dimensional embedding space suitable for channel charting. 

In this section, we introduce two CC learning functions: one based solely on TDoA extracted from single-shot CIR data, and another that combines CIR pairs with their corresponding UE displacements and TDoA.
\subsubsection{TDoA Loss Function}
We assume that the pilot signal sent by the UE is received at all TRPs and the pre-processing step is completed. First, for each $\Delta \hat\tau_{k,m_k,t}$ from a single shot of $\mathbf{H}^{\mathrm{norm}}_{t}$, we write a TDoA loss function

\begin{equation}\label{tdoa_loss_1}
    \ell^{\Delta \tau}_{k,m_k,t} = c|\Delta \tau_{k,m_k,t} - \Delta \hat\tau_{k,m_k,t}|.
\end{equation}

By incorporating the UE location $\mathbf{u}_t$ and TRP locations $\mathbf{x}_{m_k}$ we get
\begin{align}\label{loss_tdoa_2}
\ell^{\Delta \tau}_{k,m_k,t}
   &= \Big| \, \|\mathbf{x}_{m_k} - \mathbf{u}_t\|
      - \|\mathbf{x}_{\text{ref}_k} - \mathbf{u}_t\| \nonumber \\
   &\quad - c \,\Delta \hat\tau_{k,m_k,t} \,\Big|,
\end{align}
where $c$ is the constant speed of light.
As $\mathbf{u}_t$ is unknown, we equivalently replace it by the CC function $f_{\boldsymbol{\theta}}(\mathbf{H}_t^{\text{norm}})$ in \eqref{loss_tdoa_2} to have

\begin{align}\label{eq:tdoa_loss1}
\ell^{\Delta \tau}_{k,m_k,t}
   &= \Bigg| \, \big\| \mathbf{x}_{m_k} - f_{\boldsymbol{\theta}}(\mathbf{H}_t^{\text{norm}}) \big\|
      - \big\| \mathbf{x}_{\text{ref}_k} - f_{\boldsymbol{\theta}}(\mathbf{H}_t^{\text{norm}}) \big\| \nonumber \\
   &\quad  - c \, \Delta \hat{\tau}_{k,m_k,t} \,\Bigg|.
\end{align}

In realistic scenarios, LoS paths may be obstructed by obstacles or diffraction, leading to noisy TDoA estimates. Such corrupted measurements degrade the performance of the CC function. To mitigate this effect, we introduce a binary masking coefficient $\nu_{m_k,t}$ that selectively discards the NLoS-corrupted measurements. Therefore we define the final TDoA loss function as
\begin{align}\label{eq:tdoa_loss}
\mathcal{L}^{\Delta \tau} 
   &= \frac{1}{TK (M_k-1)} 
      \sum_{t=1}^{T} \sum_{k=1}^{K} \sum_{\substack{m_k=1 \\ m_k \neq m_{\text{ref}_k}}}^{M_k} \nu_{m_k,t}\ell^{\Delta \tau}_{k,m_k,t}.
\end{align}

This formulation requires a classification function which maps normalized CIR measurements to a binary LoS/NLoS masking vector
\begin{equation}
    g_{\boldsymbol{\phi}}\!\left(\mathbf{H}_t^{\text{norm}}\right) 
= \boldsymbol{\nu}_{t},
\end{equation}
where \(\boldsymbol{\nu}_t = (\nu_{1,t}, \nu_{2,t}, \dots, \nu_{M-K,t}) \in \{0,1\}^{M-K}\) is the binary masking vector whose length equals the number of available \(M-K\) TDoA measurements at time \(t\). Each element assigns a weight of 0 to discard an NLoS TDoA measurement and 1 to retain a valid LoS TDoA measurement.

In a realistic scenario, either the shortest ToA or the strongest peak power, among the antennas of one RU is taken as the reference. While this reference antenna indexing can change along the time and as the UE is in various positions. Therefore, we take a hybrid TDoA referencing by calculating every TDoA combination on an RU per timestamp, Although this is redundant in theory, hybrid referencing lowers the chance of losing useful TDoA values on an RU in some timestamps.

The details of our proposed NLoS classification function are provided later in part~\ref{sec:nlos_func} of this subsection, where we describe how 
$\boldsymbol{\nu}_t$ is derived. 
Figure~\ref{fig:tdoa_training} illustrates the training phase of our CC model using the TDoA loss function from~\eqref{eq:tdoa_loss}.

\subsubsection{TDoA+Displacement Loss Function}
As the UE moves along a trajectory, the measured data naturally exhibits temporal correlation and consecutive CIR observations are similar. This property can be leveraged by introducing a bilateral loss that enforces consistency between neighboring samples in the embedding space. Except when sudden channel variations occur due to movement or environmental changes, which will be addressed in the next part.
Consequently, the single-shot TDoA loss is extended to a bilateration TDoA loss that jointly processes pairs of measurements to preserve local geometric continuity along the UE trajectory.

We begin by randomly selecting pairs of CIR samples $(\mathbf{H}^{\text{norm}}_{t_i}, \mathbf{H}^{\text{norm}}_{t_j})$ as well as their corresponding TDoAs $(\Delta\hat{\tau}_{t_i},  \Delta\hat{\tau}_{t_j})$ from the dataset. Only pairs with timestamps $t_i$ and $t_j$ are retained if their temporal interval satisfies $|t_j - t_i| \leq \epsilon$ as defined in~\eqref{eq:disp}. Otherwise, they are discarded, and new pairs are drawn. 
Given the corresponding TDoA measurement pair, the former TDoA loss function in~\eqref{eq:tdoa_loss1} extends to
\begin{align}\label{eq:tdoa_loss_pair}
   \ell^{\Delta\tau}_{t_i,t_j} 
   &=  \nu_{m_k,t_i} \, \ell^{\Delta\tau}_{k,m_k,t_i} 
      + \nu_{m_k,t_j} \, \ell^{\Delta\tau}_{k,m_k,t_j}.
\end{align}

Moreover, in cases where the channel experiences significant changes such as LoS/NLoS transitions or when only limited number of LoS TRPs are available, we incorporate the displacement of the UE $\hat{d}_{i,j}$ between timestamps $t_i$ and $t_j$, with the TDoA CC model.
Hence, a displacement loss is defined;
\begin{equation} \label{eq:displacement_loss}
\ell^{d}_{t_i, t_j} = 
\Big| \,
\| f_{\boldsymbol{\theta}}(\mathbf{H}^{\text{norm}}_{t_i})
   - f_{\boldsymbol{\theta}}(\mathbf{H}^{\text{norm}}_{t_j}) \|
- \hat{d}_{i,j} \,\Big|.
\end{equation}

Adding the displacement term from~\eqref{eq:displacement_loss} to the TDoA loss in~\eqref{eq:tdoa_loss_pair} penalizes excessive fluctuations in the position predictions generated by the TDoA CC model along a trajectory. The overall training objective thus integrates both terms across all $T$ time steps, $K$ RUs and their $M_k$ TRPs as   
\begin{align}\label{eq:tdoa_disp_loss}
    \mathcal{L} &=
    \frac{1}{T K (M_k-1)} 
      \sum_{\substack{t_i,t_j=1 \\ |t_i - t_j| < \epsilon}}^{T} 
      \sum_{k=1}^{K} 
      \sum_{\substack{m_k=1 \\ m_k \neq m_{\text{ref}_k}}}^{M_k} \ell^{\Delta\tau}_{t_i,t_j} + \beta \, \ell^{d}_{t_i, t_j},
\end{align}
where $\beta$ controls the relative weight of the displacement term.
Figure~\ref{fig:tdoa_disp_training} illustrates the CC training process described in~\eqref{eq:tdoa_disp_loss}, where a pair of DNNs embed two corresponding CIR measurements and jointly integrate their outputs $(\hat{\mathbf{u}}_i, \hat{\mathbf{u}}_j)$ into a displacement-based loss. This loss enforces the preservation of spatial relationships by aligning similarities in the physical space with those in the learned embedding space.

\subsubsection{NLoS Classification Function}\label{sec:nlos_func}

\begin{figure}[t]
    \centering
    \begin{minipage}[t]{0.48\linewidth}
        \centering
        \includegraphics[width=\linewidth]{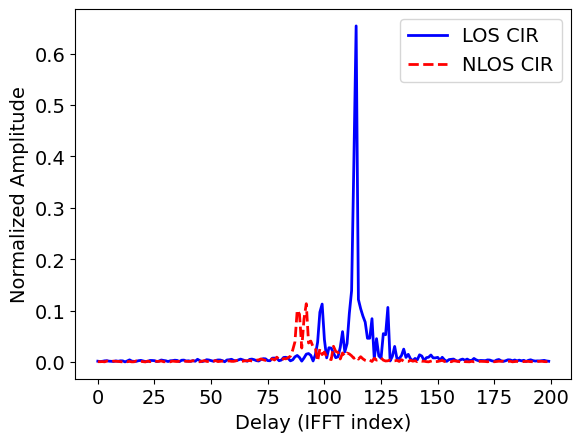}
        \caption*{(a)}
    \end{minipage}
    \hfill
    \begin{minipage}[t]{0.48\linewidth}
        \centering
        \includegraphics[width=\linewidth]{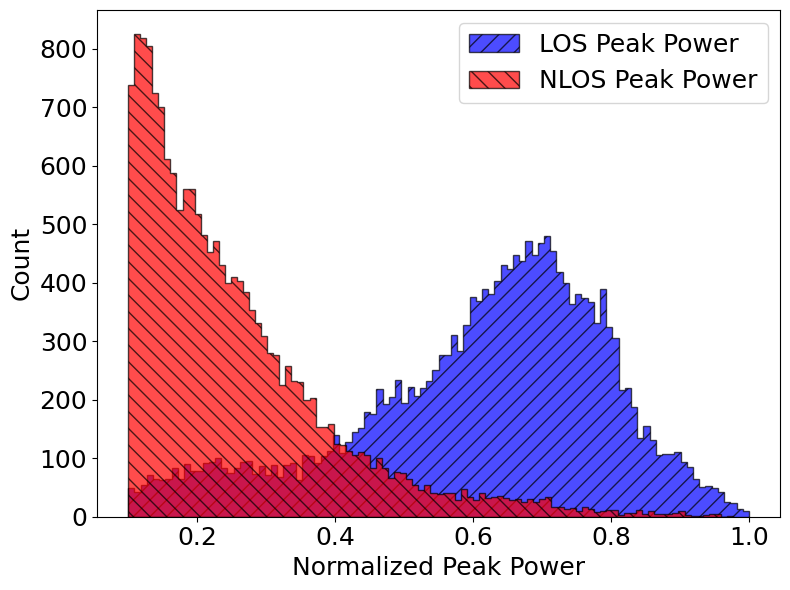}
        \caption*{(b)}
    \end{minipage}
    \caption{Sample LoS/NLoS data (a) normalized CIR traces, and (b) normalized peak power histograms}
    \label{fig:losnlos-comparison}
\end{figure}

By leveraging the empirical distributions of peak power values in large CIR datasets, we adopt a simple yet effective thresholding approach to distinguish between LoS and NLoS conditions and mask out the NLoS measurements. This method is motivated by the observations that LoS components typically exhibit significantly higher peak magnitudes than their NLoS counterparts. The contrast in their power distributions thus serves as a reliable discriminative feature, as illustrated in Figure~\ref{fig:losnlos-comparison}. 

Let us take two TRPs, $\mathbf{x}_{m_k}$ and $\mathbf{x}_{\text{ref}_k}$ involved in each TDoA measurement, then each masking element $\nu_{m,t}$ is defined as
\begin{equation}
     \nu_{m,t} =  \mu_{m_k,t} \mu_{\text{ref}_k,t}, 
\end{equation}
where $\mu_{m_k,t} \in \{0,1\}$ and $\mu_{\text{ref}_k,t} \in \{0,1\}$ are the LoS indication coefficients of the corresponding TRPs. 
%This formulation requires a classification function $g_{\boldsymbol{\phi}}(\mathbf{H}_t^{\text{norm}}) = \boldsymbol{\mu}_{t}$, where $\boldsymbol{\mu}_{t} = (\mu_{1,t}, \mu_{2,t}, \dots, \mu_{M,t}) \in \{0,1\}^M$ is a binary vector assigning a LoS/NLoS state to each TRP at time $t$.  
To compute $\mu_{m_k,t}$ we take the corresponding normalized row from $\mathbf{H}_t^{\text{norm}}$. The normalization from~\eqref{eq:normalized} drives the peak power of the NLoS rows $\mathbf{h}_{k,m_k,t}^{\text{norm}}$ close to zero. The LoS indicator is then obtained by thresholding the maximum peak of the normalized CIR:
\begin{equation}\label{eq:los_thr}
    \mu_{m_k,t} =
    \begin{cases}
        1, & \max(\mathbf{h}_{k,m_k,t}^{\text{norm}}) > \lambda \quad \text{(LoS)}, \\[6pt]
        0, & \max(\mathbf{h}_{k,m_k,t}^{\text{norm}}) \leq \lambda \quad \text{(NLoS)}.
    \end{cases}
\end{equation}

Here, $\lambda \in (0,1)$ is chosen based on the empirical distribution of normalized CIR peaks. We compute $\mu_{\text{ref}_k,t}$ similarly. This yields a lightweight and interpretable rule for LoS/NLoS classification, whose impact on CC accuracy is further examined in Section~\ref{sec:results}.

\begin{table}[t]
\centering
\small % Smaller font size for the table
\begin{tabularx}{\columnwidth}{@{}Xlll@{}}
\toprule
Layer & Output Dimension & Kernel Size & Activation \\ \midrule
Conv2D & (32, 16, 100) & (3, 3) & ReLU \\
Conv2D & (64, 16, 100) & (3, 3) & ReLU \\
Flatten & (1, 102400) & - & - \\
Fully Con. & (1, 512) & - & ReLU \\
Fully Con. & (1, 128) & - & ReLU \\
Fully Con. & (1, 2) & - & - \\ \bottomrule
\end{tabularx}
\captionsetup{size=small} % Smaller caption
\caption{DNN Embedding Model Architecture Summary}\label{tab:embedding_model_summary}
\end{table}

\subsection{Testing Channel Charting Models}
During training, our models learn to capture both TDoA and displacement information. Consequently, in the testing phase as depicted in Figure~\ref{fig:CCtesting}, the trained mapping function $f^{\ast}_{\boldsymbol{\theta}}(\cdot)$ operates solely on the pre-processed CIR inputs, without requiring explicit displacement or TDoA feature annotation.

\section{Evaluations and Results}\label{sec:results}
In this section, we evaluate the proposed models in two complementary forms: first through controlled experiments using Matlab simulations (Subsection~\ref{sec:matlabsim}), and then through real-world measurements on the GEO-5G testbed (Subsection~\ref{sec:geo5gtestbed}).

\subsection{Matlab Simulations}\label{sec:matlabsim}
To evaluate the performance of our proposed CC methods, we first develop a Matlab-based simulation environment that models the 5G positioning scenario under controlled conditions. This allows us to generate synthetic CIRs, introduce noise and multipath effects, and systematically vary parameters such as the number of antennas, bandwidth, and signal-to-noise ratio. Using this setup, we assess the accuracy, robustness, and convergence behavior of the proposed method before validating it on the real testbed.  

\begin{figure}
    \centering
    \includegraphics[width=0.99\linewidth]{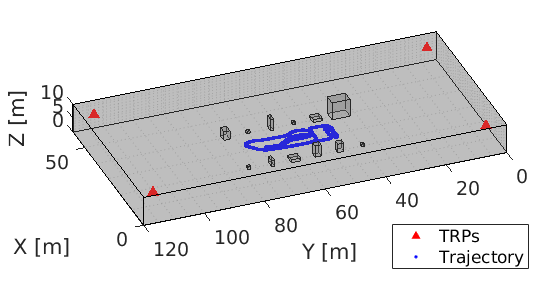}
    \caption{Simulated 3D-map InFH in Matlab environment}
    \label{fig:InFH}
\end{figure}

\begin{figure}
    \centering
    \includegraphics[width=0.85\linewidth]{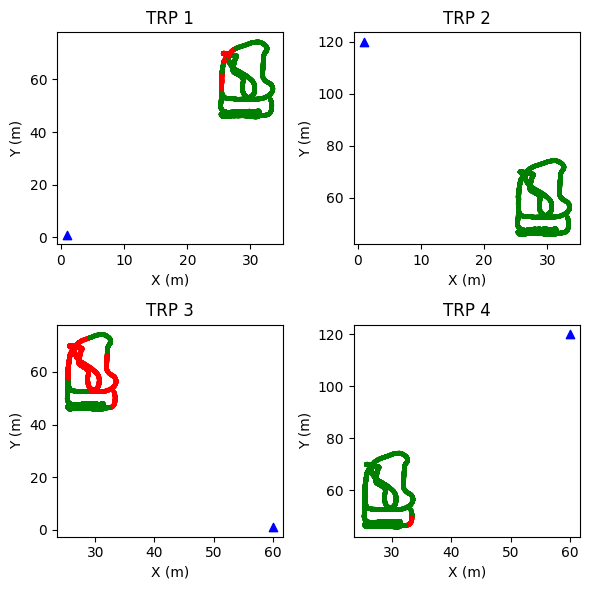}
    \caption{Simulated LoS (green) and NLoS (red) state from Each TRP's Point of View}
    \label{fig:losnlos_matlab}
\end{figure}

In this simulation, a 3D-map of an Indoor Factory Hall (InFH) environment as depicted in Figure~\ref{fig:InFH} is created, where a UE transmits UL-SRS signals with a bandwidth of $100\,\text{MHz}$ and a sampling rate of $122.88\times10^6\,\text{samples/second}$ in the sub-6\, GHz band. The signals are received by $M=4$ TRPs positioned at a fixed height of $8\,\text{m}$ that often become NLoS to the UE because of the present clutter in between. The LoS/NLoS state from each TRP's point of view during the training trajectory is illustrated in 
Figure~\ref{fig:losnlos_matlab}. Based on this trajectory, the CIR data as well as the displacement data are reconstructed in MATLAB using a 3D map-based Ray-Tracing toolbox. 

\begin{figure}
    \centering
    \includegraphics[width=0.9\linewidth]{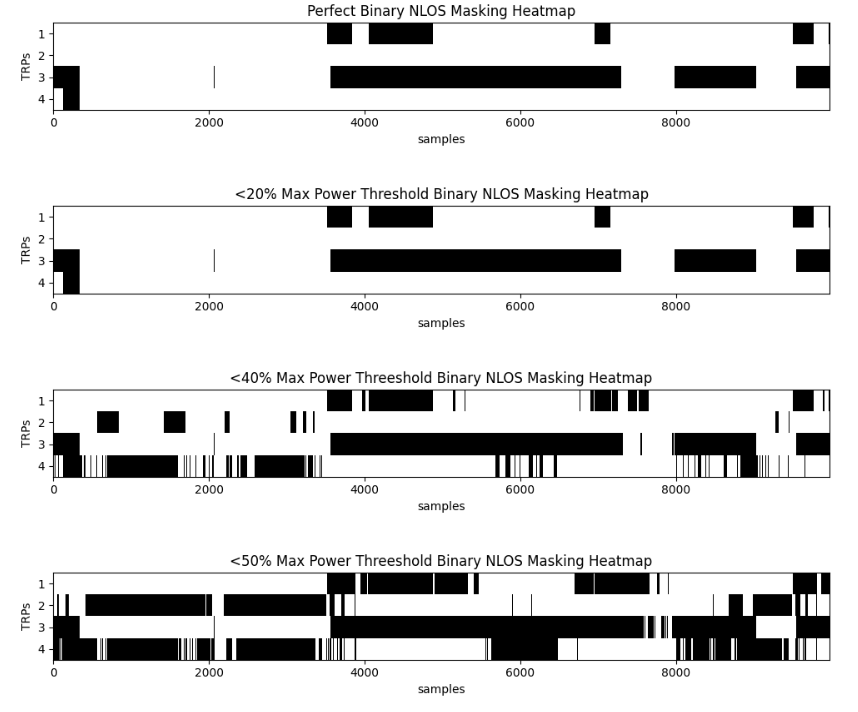}
    \caption{Simulated NLoS masking heatmaps (TRPs vs. samples): ground truth and predicted masks for $\lambda=0.2$, $0.4$, and $0.5$ (black = masked/NLoS, white = kept/LoS) illustrating that larger $\lambda$ values increasingly overestimate NLoS regions.}
    \label{fig:heatmap_matlab}
\end{figure}

We first examine the effectiveness of masking in a mixed LoS/NLoS scenario by applying different threshold values and comparing them to the case without masking.  
Figure~\ref{fig:heatmap_matlab} shows the heatmap of NLoS masking for a number of threshold values $\lambda$. The results show that $\lambda = 0.2$ provides the closest match to the perfect NLoS identification, while higher thresholds ($\lambda = 0.4$ or $\lambda = 0.5$) tend to overestimate the NLoS regions. Consequently, where only 4 TRPs are available, some valid LoS TDoAs are incorrectly discarded, which reduces the number of usable TRP pairs and makes the localization problem under-constrained, leading to worse accuracy than the no-masking case.

\begin{figure}
    \centering
    \includegraphics[width=0.9\linewidth]{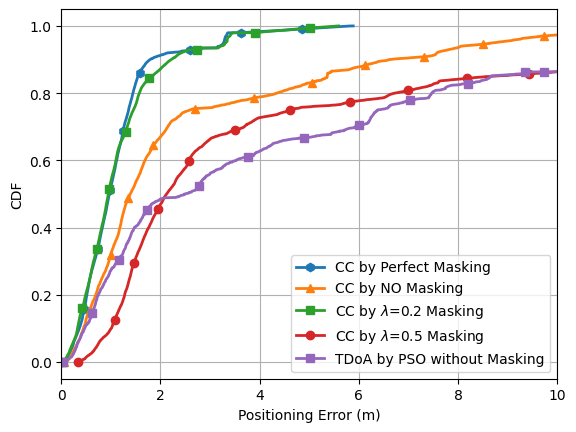}
    \caption{Simulated positioning Error CDF of TDoA CC with NLoS Masking}
    \label{fig:cc_tdoa_matlab}
\end{figure}

The positioning performance of the CC method with TDoA under different values of $\lambda$ is presented in Figure~\ref{fig:cc_tdoa_matlab}. For comparison, the result of a conventional TDoA approach using Particle Swarm Optimization (PSO)~\cite{kennedy1995particle} is also included. 
Our proposed TDoA-based CC model with NLoS masking at $\lambda = 0.2$ achieves performance comparable to the perfect masking case, reaching a localization accuracy of approximately $2.5,\text{m}$ at the 90\% of the CDF.
In contrast, when no masking is applied or a high threshold (e.g., $\lambda = 0.5$) is used, the performance of the CC model degrades noticeably.
Moreover, the TDoA-based PSO without masking exhibits the largest positioning errors, primarily due to its lack of robustness against NLoS conditions. This limitation is also inherent in conventional 3GPP positioning systems employing a Location Management Function (LMF) like in Figure~\ref{fig:cc_systemmodel}, where the positioning protocols (such as NRPPa and LPP) do not provide access to extended CSI data that could enable effective NLoS mitigation.

The next hyper-parameter that balances the contribution of the displacement loss relative to the TDoA loss is $\beta$ in~\eqref{eq:tdoa_disp_loss}. By exploring several values, empirically, $\beta = 2$ was found to align the numerical scales of both error terms and to yield the minimum prediction error. This value can, however, be adjusted depending on the characteristics and precision of the velocity or displacement sensor used in the system. 

\begin{figure}
    \centering
    \includegraphics[width=0.9\linewidth]{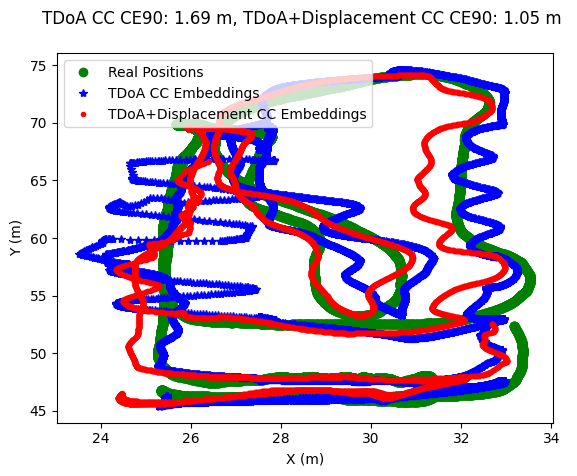}
    \caption{Simulated CC Embeddings with NLoS Masking fixed at $\lambda=0.2$ and $\beta=2$}
    \label{fig:CC_tdoa_disp_matlab}
\end{figure}

\begin{figure}
    \centering
    \includegraphics[width=0.9\linewidth]{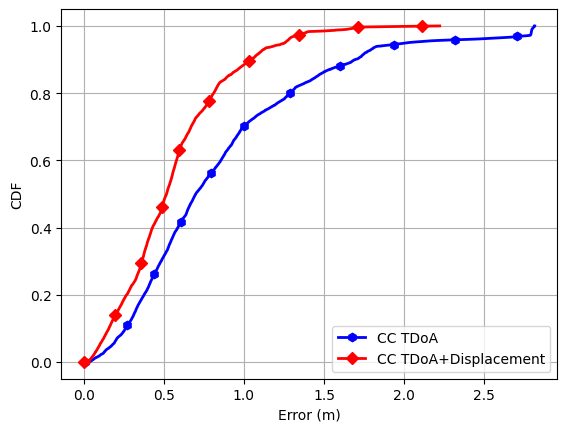}
    \caption{Simulated CDF NLoS Masking with fixed $\lambda=0.2$ and $\beta=2$}
    \label{fig:cdf_tdoa_disp_matlab}
\end{figure}

The inference results using the previously fixed model parameters are presented in Figures~\ref{fig:CC_tdoa_disp_matlab} and~\ref{fig:cdf_tdoa_disp_matlab}, showing the estimated trajectories and error CDFs for both the TDoA-only CC model and the joint TDoA+Displacement CC model.

%\begin{figure}
%    \centering
%    \includegraphics[width=0.85\linewidth]{figures/results2%/traj.png}
%    \caption{Low TRP LoS/NLoS state from Each TRP's Point of View}
%    \label{fig:enter-label}
%\end{figure}

%\begin{figure}
%    \centering
%    \includegraphics[width=0.85\linewidth]{figures/results2/masking_heatmap.png}
%    \caption{Low TRP NLoS Masking Heatmap}
%    \label{fig:enter-label}
%\end{figure}

%\begin{figure}
%    \centering
%    \includegraphics[width=0.85\linewidth]{figures/results2/cdf_masking.png}
%    \caption{Low TRP Positioning Error CDF by LoS Masking}
%    \label{fig:enter-label}
%\end{figure}

%\begin{figure}
%    \centering
%    \includegraphics[width=0.85\linewidth]{figures/results2/embeddings.png}
%    \caption{Low TRP Embeddings with NLoS Masking}
%    \label{fig:enter-label}
%\end{figure}

\subsection{GEO-5G Testbed}\label{sec:geo5gtestbed}
\begin{figure*}[t]
    \centering
    \includegraphics[width=0.9\textwidth]{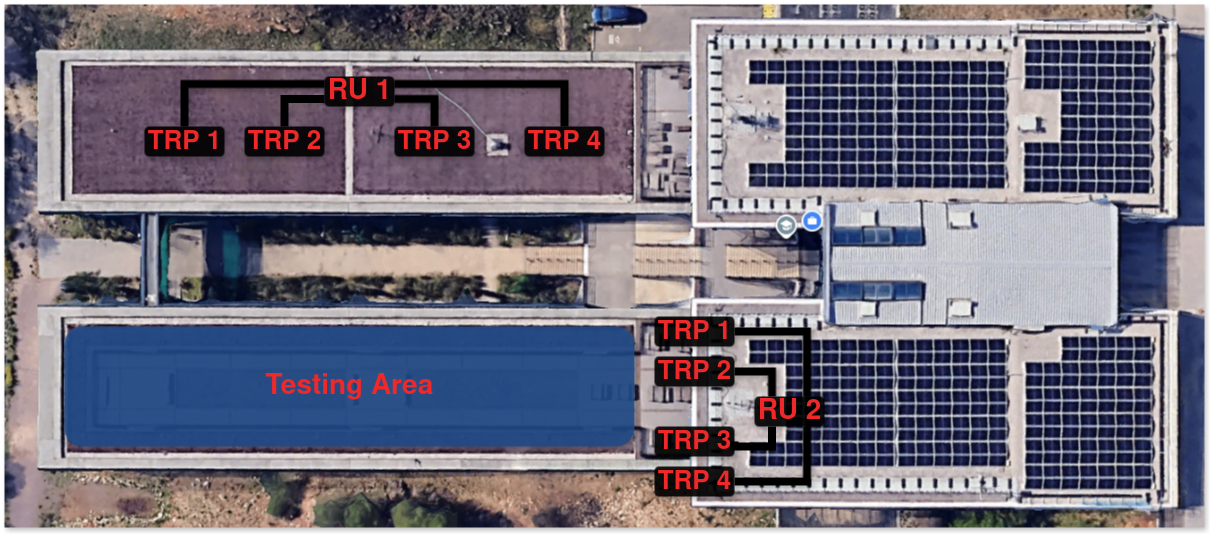}
    \caption{GEO-5G testbed: Deployment of 2 O-RUs each with 4 distributed TRPs on the south terrace and the north rooftop of EURECOM, with a testing area of size 50x10 on the north terrace}
    \label{fig:testbedmap}
\end{figure*}
The GEO-5G localization testbed is part of EURECOM's Open5G testbed, featuring high-speed fiber-connected computing and switching infrastructure. It supports virtualized 5G deployment with USRP and O-RAN radios, integrating OAI for virtualized network functions.
Leveraging our recent contributions to OAI RAN reported in~\cite{malik2024concept}, we can choose between a single-gNB with multi-RUs or a multi-gNB with multi-RUs architecture.
To evaluate OAI's new localization features, we deployed two VVDN O-RAN RUs~\cite{vvdn_oran_portfolio_2023}, provided by Firecell~\cite{firecell_5g}
with a single gNB (CU-DU) and integrated them into the EURECOM 5G testbed. Furthermore, we have four distributed Panorama directional antennas~\cite{panorama_antennas} as TRPs on each RU, which are mounted on the roof railings with low-loss cables, which gives us a total of eight TRPs to cover a testing area of 50m$\times$10m on the north terrace of the EURECOM building, see Figure~\ref{fig:testbedmap}. 
While the integration of the E2 interface and RIC was still in progress during the preparation of this paper, the performance evaluation was conducted using the MQTT protocol instead of E2. CIR data were transferred from the CU-DU to an AI/ML host machine via MQTT to enable real-time prediction of UE positions. At the same time, the host machine collects RTK position measurements and their corresponding timestamps from the RTK rover over Wi-Fi. These measurements are then used to evaluate the real-time error of the model’s predictions, with RTK measurements serving as the ground truth.

In the remainder of this section, we outline the key practical considerations for implementing and operating our 5G positioning testbed, followed by a presentation of its performance evaluation and results.

\subsubsection{Synchronization} \label{sec:synch}
To achieve precise synchronization among distributed RUs and DU, we employ the IEEE 1588-based PTP over the Fronthaul Synchronization Plane (S-Plane) via Ethernet between the DU and RU~\cite{oran_splane}. A Qulsar Qg2 Grandmaster Clock serves as the central timing source, which locks its timebase to GNSS (Global Navigation Satellite System) signals, ensuring nanosecond-level absolute time accuracy. The Grandmaster then distributes this timing information over fiber to all RUs via a PTP-compatible switch. Each RU operates as a PTP worker, adjusting its local oscillator to align with the Grandmaster's reference clock. This synchronization is critical for coherent UL-TDoA measurements. 

\begin{figure}
    \centering
    \includegraphics[width=1\linewidth]{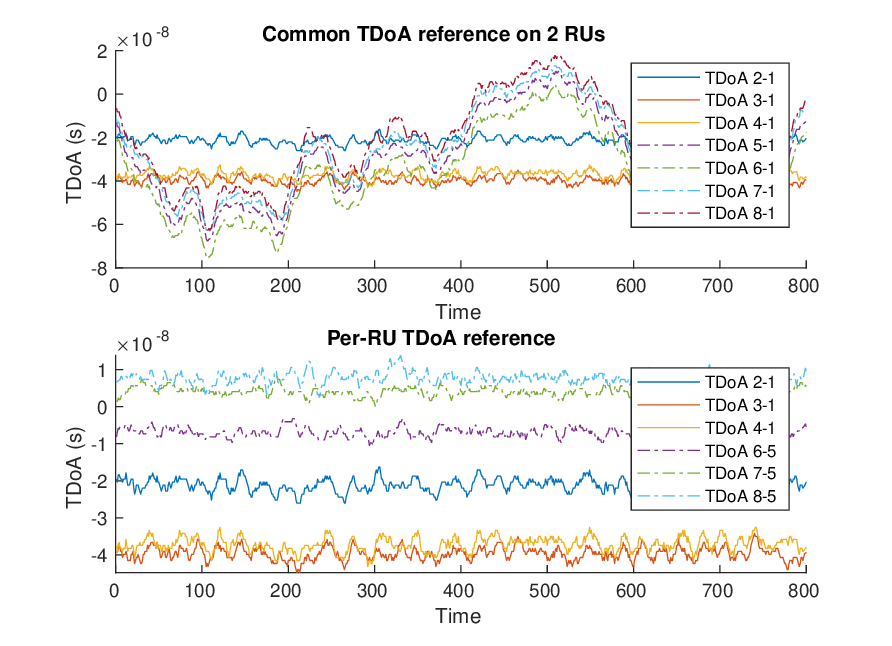}
    \caption{Common and Per-RU TDoA referencing approaches under synchronization errors among RUs with a stationary UE}
    \label{fig:per_ru_tdoa}
\end{figure}

However, due to PTP switch impairments and RU long distances at this testbed, errors up to 40 ns are observed. To mitigate the propagation of clock drift from one RU to the other, we adopted a per-RU reference TRP strategy in calculating TDoAs instead of a common reference across all RUs, ensuring internal consistency and reducing cross-RU time drift. These modifications alter the classical TDoA-based positioning. Therefore, a conventional TDoA-based positioning algorithm suffers from these impairments. Also, a tailored pre-processing pipeline in Figure~\ref{fig:preprocessing_diagram} is employed to discard outlier measurements. Figure~\ref{fig:per_ru_tdoa} illustrates the impact of RU-to-RU timing drift on the GEO-5G testbed using a stationary UE (2 RUs, 4 TRPs each). With a common TDoA reference across both RUs (top), the TDoAs of the second RU exhibit large fluctuations due to inter-RU clock offsets. Using a per-RU reference (bottom) removes this inter-RU bias and preserves stability with cost of a few less effective TDoA measurements.

\subsubsection{Error Measurement} \label{sec:GT}
RTK enhances conventional GPS measurements by resolving the carrier-phase ambiguities between a fixed \textit{Base} module and a mobile \textit{Rover} module, enabling a promising centimeter-level accuracy in localization.
Therefore, we employ RTK to benchmark our positioning system and use its data to calculate the error of our CC predictions. Although RTK provides accurate absolute positions in geographic coordinates (latitude, longitude, altitude), our system operates in a relative Cartesian reference frame. To bridge this, we transformed the geographic coordinates into local $[u_x, u_y, u_z]$ Cartesian coordinates by applying appropriate projection and translation techniques.
Open-source geospatial libraries such as \textit{pymap3d} in Python facilitate the conversion of geographic coordinates into raw Cartesian coordinates (x-East, y-North, z-Up) known as ENU, through standard projection methods. 

Furthermore, to align these raw Cartesian coordinates with our system's local reference frame (where TRP 1 on RU 1 is (0, 0, 2.2)), we perform an additional linear transformation. Specifically, we compute an affine transformation matrix based on a set of predefined anchor points with known correspondences in both geographic and local coordinate systems. This matrix is then used to map all future positions consistently into the local reference frame, ensuring spatial alignment with the experimental environment.
We align the raw Cartesian coordinates \(\mathbf{u}_{\text{raw}} \in \mathbb{R}^3\), converted from geographic coordinates, with our local coordinate system \(\mathbf{u}_{\text{local}} \in \mathbb{R}^3\) using an affine transformation of the form:

\begin{equation}
    \mathbf{u}_{\text{local}} = \mathbf{A} \cdot \mathbf{u}_{\text{raw}} + \mathbf{b}
\end{equation}
where \(\mathbf{A} \in \mathbb{R}^{3 \times 3}\) is a linear transformation matrix, and \(\mathbf{b} \in \mathbb{R}^3\) is a translation vector. $\mathbf{A}$ and $\mathbf{b}$ are estimated simply by a few known local and raw points using the Linear Least Squares estimator and recorded to be used in real-time testing.
\begin{comment}
Given a set of \(N\) known corresponding point pairs \(\left\{ \left( \mathbf{x}_{\text{raw}}^{(i)}, \mathbf{x}_{\text{local}}^{(i)} \right) \right\}_{i=1}^N\), we estimate \(\mathbf{A}\) and \(\mathbf{b}\) using least squares. We first construct the matrices:

\[
\mathbf{X}_{\text{raw}} =
\begin{bmatrix}
(\mathbf{x}_{\text{raw}}^{(1)})^\top \\
\vdots \\
(\mathbf{x}_{\text{raw}}^{(N)})^\top
\end{bmatrix}
\in \mathbb{R}^{N \times 3}, \quad
\mathbf{X}_{\text{local}} =
\begin{bmatrix}
(\mathbf{x}_{\text{local}}^{(1)})^\top \\
\vdots \\
(\mathbf{x}_{\text{local}}^{(N)})^\top
\end{bmatrix}
\in \mathbb{R}^{N \times 3}
\]

To estimate both \(\mathbf{A}\) and \(\mathbf{b}\) simultaneously, we augment \(\mathbf{X}_{\text{raw}}\) with a column of ones:

\[
\tilde{\mathbf{X}}_{\text{raw}} =
\begin{bmatrix}
\mathbf{x}_{\text{raw}}^{(1)} & 1 \\
\vdots & \vdots \\
\mathbf{x}_{\text{raw}}^{(N)} & 1
\end{bmatrix}
\in \mathbb{R}^{N \times 4}
\]

We then solve the least squares problem:

\[
\begin{bmatrix}
\mathbf{A} & \mathbf{b}
\end{bmatrix}^\top = 
\left( \tilde{\mathbf{X}}_{\text{raw}}^\top \tilde{\mathbf{X}}_{\text{raw}} \right)^{-1}
\tilde{\mathbf{X}}_{\text{raw}}^\top \mathbf{X}_{\text{local}}
\]

This yields a single affine transformation that maps any future \(\mathbf{x}_{\text{raw}}\) to the local coordinate frame:

\[
\mathbf{x}_{\text{local}} = 
\begin{bmatrix}
\mathbf{A} & \mathbf{b}
\end{bmatrix}
\begin{bmatrix}
\mathbf{x}_{\text{raw}} \\
1
\end{bmatrix}
\]
\end{comment}

\subsubsection{Spatial Power Profile}
\begin{figure}[htbp]
    \centering
    \includegraphics[width=1\linewidth]{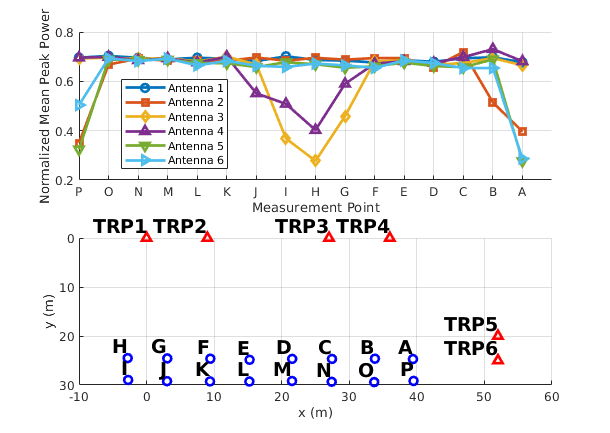}
    \caption{GEO-5G tesbed: TRPs spatial power profile}
    \label{fig:powerdecay}
\end{figure}

There are three main sources of power decay in the SRS received signal and consequently channel gain. Distance, multipath, and TRP's beam pattern.
While in theory we expect the received power to be attenuated with the larger distances of UE, the UL power control in real 5G systems plays an essential role in providing a steady channel gain between transmitter and receiver. Specifically, OAI software makes this power control on port 1 of each RUs. Figure~\ref{fig:powerdecay} validates this on TRPs 1 and 5, who experience steady normalized peak power as the UE is placed at various distances to them. 

However, a NLoS path can show a significantly lower received power than we expect. This is observed on points A and P that are under the building and NLoS to TRP 5 and 6, where the normalized mean peak power is significantly lower.

Moreover, each TRP has a certain angle coverage in azimuth and elevation. The 3dB beamwidth or Half Power Beam Width (HPBW) is the angular width of the main lobe between the two directions where the gain falls by 3 dB, or equivalently, where the power is half of the maximum value. In this testbed, we deployed TRPs with $50^{\circ}$ beamwidth in (Azimuth, Elevation). This effect can be observed when the UE is located near the center of the testing area, around points BCDEF and KLMNO, where the TRPs receive a steady and higher normalized peak power. In contrast, TRPs 2, 5, and 6 experience reduced power when the UE is positioned at points ABPO, and a similar reduction is seen for TRPs 3 and 4 when the UE is at points GHIJ, which lie outside of their beam coverage.

\subsubsection{Synthetic NLoS Measurements}
The dataset employed in this work was collected using a handheld mobile phone, which was subject to undesired self-shadowing effects caused by the person carrying it. Since no additional clutter existed between the user and the TRPs in the measurement environment, we refer to it as a LoS dataset for generality. 

\begin{figure}[htbp]
    \centering
    \includegraphics[width=0.99\linewidth]{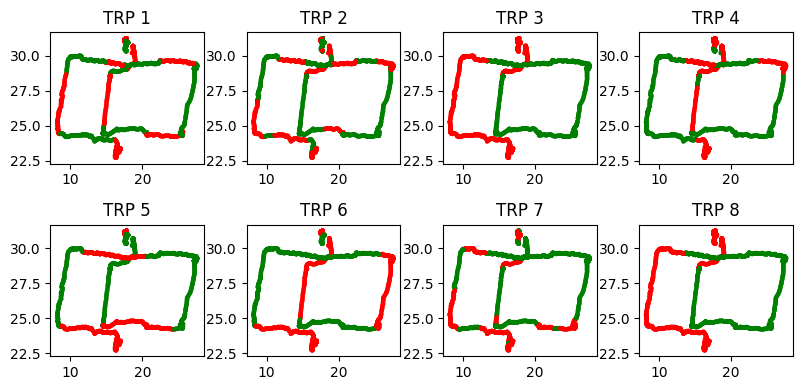}
    \caption{GEO-5G testbed: LoS (green) and NLoS (red) state of TRPs at GEO-5G testbed with $r_{\text{LoS}}=50\%$}
    \label{fig:los_nlos_testbed}
\end{figure}

Specifically, for a randomly selected subset of TRP measurements, we (i) apply a random attenuation factor to the CIR magnitude to reproduce the reduced received power typically observed under blockage, and (ii) introduce a random shift of the dominant CIR peak to emulate the timing distortion caused by multipath propagation.
This way, we evaluate the performance of the CC TDoA and TDoA+Displacement losses in different LoS ratios denoted with $r_{\text{LoS}=\{25\%, 50\%,75\%, 100\%\}}$. 
For instance, $r_{\text{LoS}}=50\%$ means $50\%$ of the measurements during the whole trajectory on all TRPs are LoS, and the rest of the trajectory are synthetically behaved as NLoS by random attenuations and random peak shiftings. See Figure~\ref{fig:los_nlos_testbed}.

\subsubsection{Displacement Measurements}
\begin{figure}[htbp]
    \centering
\includegraphics[width=0.9\linewidth]{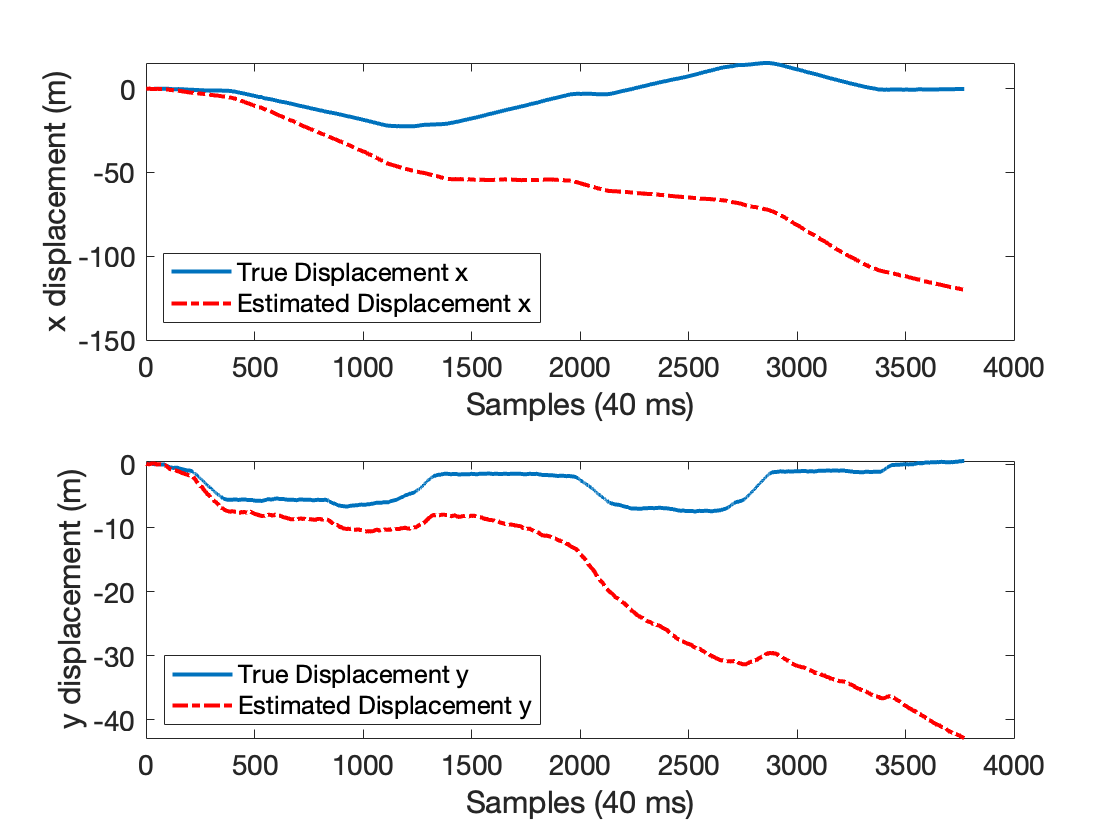}
    \caption{Displacement bias growing over time}
    \label{fig:disp_bias}
\end{figure}

Due to the absence of sensors (such as IMU, Laser, LiDAR, etc.) to provide velocity or displacement measurements in our testbed, we generated RTK-assisted synthetic displacement data. 
To avoid the influence of RTK accuracy on the synthetic displacement measurements, additive noise and growing bias were applied to the displacements over longer intervals as detailed in \eqref{eq:disp} of Section~\ref{sec:sysmodel}. 
Therefore, the displacement measurements were estimated in maximum intervals of $\epsilon=4$ seconds, without explicitly incorporating the RTK positions into the loss function. It is important to note that relying solely on displacement measurements leads to inaccurate trajectories, as noise and bias accumulate over time, shown in Figure~\ref{fig:disp_bias}. These measurements are thus reliable only over short intervals. While developing an accurate velocity or displacement estimation algorithm is beyond the scope of this paper, integrating such sensors is considered for future work.

\subsubsection{Hyperparameter selection}
The proposed framework includes three main hyperparameters: the NLoS masking threshold $\lambda$ in~\eqref{eq:los_thr}, the displacement weight $\beta$ in~\eqref{eq:tdoa_disp_loss}, and the maximum displacement interval $\epsilon$ of the measurements in~\eqref{eq:disp}. The threshold $\lambda$ is selected based on the empirical distribution of normalized CIR peak magnitudes (Figure~\ref{fig:losnlos-comparison}). In practice, values in the range $\lambda \in [0.2, 0.4]$ provide a good trade-off between discarding NLoS-corrupted measurements and retaining sufficient valid TDoA constraints. This is further validated through the masking sensitivity analysis in Figures~\ref{fig:heatmap_matlab} and~\ref{fig:cc_tdoa_matlab}. The weight $\beta$ is chosen such that the displacement loss and the TDoA loss have comparable numerical scales during training, and we found $\beta=2$ to provide stable convergence and the lowest positioning error in our experiments. Finally, $\epsilon$ is set based on the expected drift of displacement measurements; in our setup we use $\epsilon=4$ seconds to ensure displacement constraints remain reliable over short intervals while limiting accumulated bias.

\subsubsection{Tetstbed Results}
For evaluating our CC models performance at the GEO-5G testbed, we adopt the standard set of metrics widely used in the literature~\cite{10052099}. Continuity (CT) and Trustworthiness (TW) assess how well the embedding preserves local spatial relationships, while Kruskal Stress (KS) quantifies global distortion by comparing distances in the embedded space with those in the original domain. Lower KS indicates a more faithful embedding. To measure positioning accuracy, we report both the 90th-percentile localization error (CE90) and the Mean Absolute Error (MAE), which together capture the reliability and spread of localization errors. Finally, qualitative validation is performed by visually comparing the learned embeddings with the RTK-estimated trajectories, ensuring that the models retain meaningful geometrical structures.

For the experimental evaluation of our proposed models in a real-world channel, we train and test the CC models using data collected from this testbed and report the quantitative and qualitative performance of our two CC models: the TDoA-based CC model and the TDoA+Displacement CC model. 

The results in Table~\ref{tab:cc_metrics_comparison} show that the TDoA+Displacement CC model consistently outperforms the TDoA-only model across all LoS ratios. At 100\% LoS, it achieves the best performance with the highest continuity (CT = 0.9838), trustworthiness (TW = 0.9796), and the lowest localization errors (CE90 = 1.80 m, MAE = 1.24 m).
Even under degraded conditions with 50\% or 25\% LoS, the TDoA+Displacement model maintains a clear advantage, reducing CE90 by up to 1.3 m compared to TDoA-only.
Overall, incorporating displacement constraints improves both the geometrical consistency and the accuracy of CC predictions, particularly in challenging low-LoS environments.

\begin{figure}[htbp]
    \centering
    \includegraphics[width=0.99\linewidth]{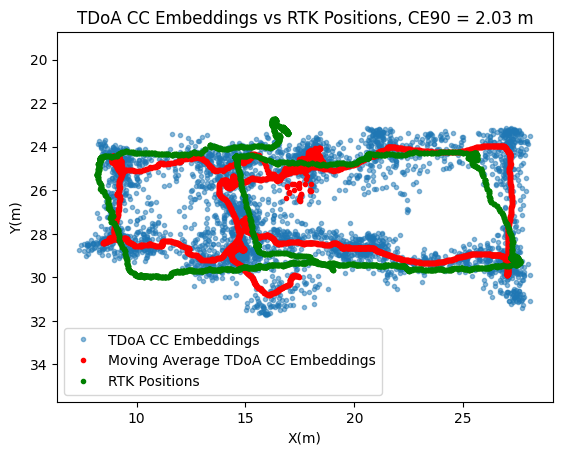}
    \caption{GEO-5G testbed: LoS TDoA CC Embeddings with $r_{\text{LoS}}=100\%$}
    \label{fig:tdoa_cc_testbed}
\end{figure}

\begin{figure}[htbp]
    \centering
    \includegraphics[width=0.99\linewidth]{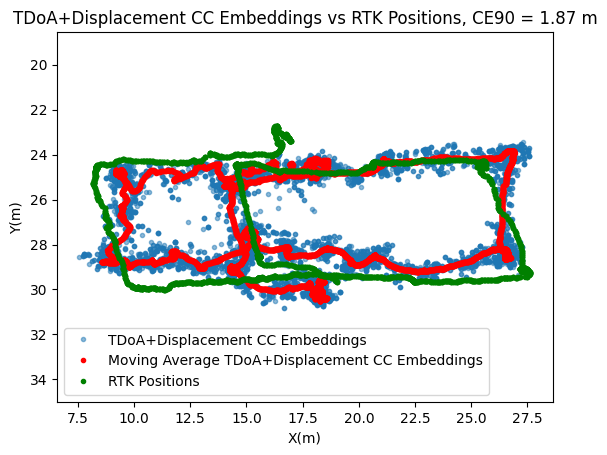}
    \caption{GEO-5G testbed: TDoA+Displacement CC Embeddings with $r_{\text{LoS}}=100\%$}
    \label{fig:joint_cc_testbed}
\end{figure}

The benefit of including displacement information can also be seen by comparing Figure~\ref{fig:tdoa_cc_testbed} and Figure~\ref{fig:joint_cc_testbed}, which show the performance of the TDoA CC method and the joint TDoA+Displacement CC method, respectively. While the TDoA CC achieves acceptable results with about $2~\text{m}$ accuracy in 90\% of the time, the joint method improves this to $1.8~\text{m}$, where both evaluated with the original LoS condition. The joint approach also yields more stable results, with reduced fluctuations even in the raw embeddings. To further enhance stability, the raw embeddings of both methods, shown in blue dots, are smoothed using a moving-average window of size 10 that is shown with red dots on the figures, while the green dots are the RTK ground truth positions.

\begin{figure}[htbp]
    \centering    \includegraphics[width=0.85\linewidth]{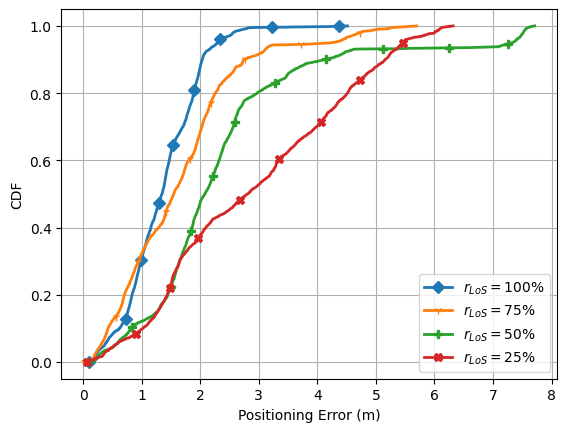}
    \caption{GEO-5G testbed: TDoA CC error CDF at GEO-5G testbed}
    \label{fig:tdoa_cc_testbed_cdf}
\end{figure}

Finally, Figure~\ref{fig:tdoa_cc_testbed_cdf} and Figure~\ref{fig:tdoa_disp_cc_testbed_cdf} compare the positioning error CDFs for the TDoA-only CC model and the proposed TDoA+Displacement CC model under different LoS availability ratios. Regarding computational complexity, classical UL-TDoA positioning typically involves NRPPa/LPP signaling overhead and iterative solvers, which can lead to end-to-end latencies on the order of hundreds of milliseconds depending on the implementation. In contrast, once trained, our DNN-based approach performs inference through a single forward pass and can output a position estimate within a few tens of milliseconds, making it more suitable for near-real-time, and smoother UE tracking. This is further analysed in our work on O-RAN framework for Near-RealTime RIC localization using CC in~\cite{bouknana2026oranframeworkaimlbasedlocalization}.

With 100\% LoS ratio, both models achieve comparable performance, around 2 m errors in most cases. However, as the LoS ratio decreases, the TDoA+Displacement model shows clear improvements: at 75\% and 50\% LoS, it reduces error spread by about 0.5–1 m, and at 25\% LoS it remains more stable than the TDoA-only baseline. These results confirm that the proposed TDoA CC method outperforms the conventional TDoA-based and state-of-the-art self-supervised methods in a realistic 5G system. Moreover, while both models perform similarly in fully LoS scenarios, the displacement-enhanced approach, combined with the NLoS masking strategy, significantly improves robustness under reduced LoS availability, consistently lowering localization errors across all partial LoS ratios.

\begin{table}[htbp]
\centering
\begin{tabular}{c c c c c c}
\hline
\textbf{$\%r_{\text{LoS}}$} & \textbf{CT} & \textbf{TW} & \textbf{KS} & \textbf{CE90 (m)} & \textbf{MAE (m)} \\
\hline
\multirow{2}{*}{100} & 0.9751 & 0.9722 & 0.1762 & 2.04 & 1.35 \\ 
                     & \textbf{0.9838} & \textbf{0.9796} & \textbf{0.1569} & \textbf{1.80} & \textbf{1.24} \\
\hline
\multirow{2}{*}{75}  & 0.9699 & 0.9529 & 0.2216 & 2.75 & 1.64 \\ 
                     & \textbf{0.9796} & \textbf{0.9722} & \textbf{0.2145} & \textbf{1.92} & \textbf{1.42} \\
\hline
\multirow{2}{*}{50}  & 0.9389 & 0.9312 & 0.2666 & 4.12 & 2.42 \\ 
                     & \textbf{0.9717} & \textbf{0.9442} & \textbf{0.2525} & \textbf{3.50} & \textbf{1.70} \\
\hline
\multirow{2}{*}{25}  & 0.9365 & 0.8655 & 0.3502 & 5.16 & 2.91 \\ 
                     & \textbf{0.9595} & \textbf{0.8966} & \textbf{0.3187} & \textbf{3.86} & \textbf{2.60} \\
\hline
\end{tabular}
\caption{GEO-5G testbed: Comparison of TDoA and \textbf{TDoA+Displacement} CC 
metrics while RTK is assumed to be the true position}\label{tab:cc_metrics_comparison}
\end{table}


\begin{figure}[htbp]
    \centering
    \includegraphics[width=0.85\linewidth]{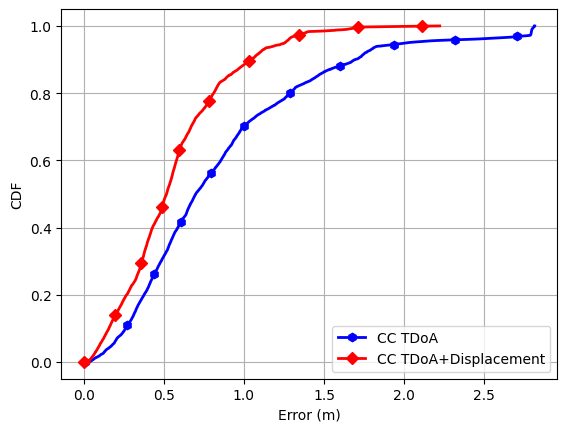}
    \caption{GEO-5G testbed: TDoA+Displacement CC error CDF at GEO-5G testbed}
    \label{fig:tdoa_disp_cc_testbed_cdf}
\end{figure}

\begin{figure}[htbp]
    \centering
    \includegraphics[width=0.85\linewidth]{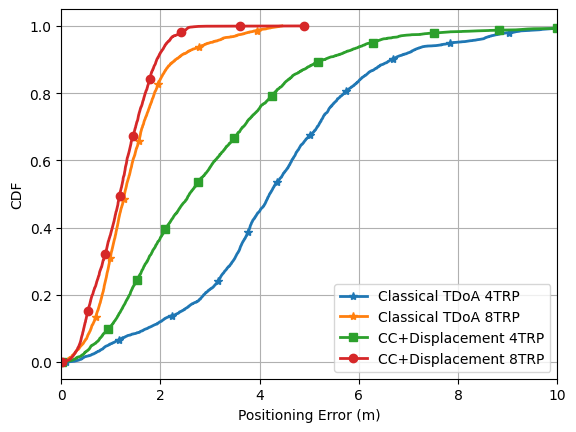}
    \caption{GEO-5G testbed: TDoA+Displacement CC error and PSO by TDoA error}
    \label{fig:tdoa_disp_cc_testbed_cdf}
\end{figure}

\section{Conclusions and Future Work}\label{sec:conclusion}
In this work, we presented a TDoA-based self-supervised CC model for UE positioning in mixed LoS/NLoS conditions. The proposed method does not require ground-truth labels for training or any coordinate scaling, and operates in a global reference frame by exploiting only the known TRP locations, TDoA, and CIR data. 

The proposed approach assumes known TRP locations and nanosecond-level precision of intra-RU synchronization for reliable UL-TDoA extraction. Moreover, the displacement constraint is only reliable over short intervals due to drift, and the masking threshold $\lambda$ may require re-tuning across deployments with different propagation conditions.

This model's performance is further improved in scenarios with limited LoS TRP availability by incorporating 
UE displacement measurements. The model was evaluated both in MATLAB simulations and in a real world GEO-5G positioning testbed at EURECOM using commercial O-RAN RUs and OAI software. 

Results of predicted positions from our joint TDoA+Displacement model compared to RTK positioning, show an error of $2$–$4\,\text{m}$ in $90\%$ of the time, across scenarios with $100\%$ to $25\%$ LoS ratios. 

The dataset of recorded CIRs, as well as RTK measurements to be used as ground truth positions, is publicly available for further research. As part of our future work, we plan to complete the integration of the E2 interface and conduct a performance evaluation of the proposed positioning system on the Near-Real-Time RIC based on OAI.

\section*{Acknowledgment}
This work has been funded by the France 2030 investment
program through the 5G-OPERA, STIC-5G and GEO-5G projects. Part of this work by Omid Esrafilian was funded by the COBOTPLANET project under the NGI Sargasso funding program.

\bibliographystyle{IEEEtran}
\bibliography{ref}
\newpage

\vfill\pagebreak

\end{document}